\documentclass[useAMS,usenatbib]{mn2e}
\usepackage{graphicx,natbib,times,amssymb}
 \title[SN Ia Time Delays]{How old are SN Ia Progenitor Systems? New Observational Constraints on the Distribution of Time Delays from \textit{GALEX}} \author[Kevin Schawinski]{
  \parbox[t]{16cm}{Kevin~Schawinski$^{1,2}$\thanks{E-mail: kevin.schawinski@yale.edu}}\\
  $^{1}$Department of Physics, Yale University, New Haven, CT 06511, U.S.A.\\
  $^{2}$Yale Center for Astronomy and Astrophysics, Yale University, P.O. Box 208121, New Haven, CT 06520, U.S.A.\\
}

\begin{document}

\newcommand\aj{{AJ}}%
\newcommand\actaa{{Acta Astron.}}%
\newcommand\araa{{ARA\&A}}%
\newcommand\apj{{ApJ}}%
\newcommand\apjl{{ApJ}}%
\newcommand\apjs{{ApJS}}%
\newcommand\ao{{Appl.~Opt.}}%
\newcommand\apss{{Ap\&SS}}%
\newcommand\aap{{A\&A}}%
\newcommand\aapr{{A\&A~Rev.}}%
\newcommand\aaps{{A\&AS}}%
\newcommand\azh{{AZh}}%
\newcommand\baas{{BAAS}}%
\newcommand\caa{{Chinese Astron. Astrophys.}}%
\newcommand\cjaa{{Chinese J. Astron. Astrophys.}}%
\newcommand\icarus{{Icarus}}%
\newcommand\jcap{{J. Cosmology Astropart. Phys.}}%
\newcommand\jrasc{{JRASC}}%
\newcommand\memras{{MmRAS}}%
\newcommand\mnras{{MNRAS}}%
\newcommand\na{{New A}}%
\newcommand\nar{{New A Rev.}}%
\newcommand\pra{{Phys.~Rev.~A}}%
\newcommand\prb{{Phys.~Rev.~B}}%
\newcommand\prc{{Phys.~Rev.~C}}%
\newcommand\prd{{Phys.~Rev.~D}}%
\newcommand\pre{{Phys.~Rev.~E}}%
\newcommand\prl{{Phys.~Rev.~Lett.}}%
\newcommand\pasa{{PASA}}%
\newcommand\pasp{{PASP}}%
\newcommand\pasj{{PASJ}}%
\newcommand\qjras{{QJRAS}}%
\newcommand\rmxaa{{Rev. Mexicana Astron. Astrofis.}}%
\newcommand\skytel{{S\&T}}%
\newcommand\solphys{{Sol.~Phys.}}%
\newcommand\sovast{{Soviet~Ast.}}%
\newcommand\ssr{{Space~Sci.~Rev.}}%
\newcommand\zap{{ZAp}}%
\newcommand\nat{{Nature}}%
\newcommand\iaucirc{{IAU~Circ.}}%
\newcommand\aplett{{Astrophys.~Lett.}}%
\newcommand\apspr{{Astrophys.~Space~Phys.~Res.}}%
\newcommand\bain{{Bull.~Astron.~Inst.~Netherlands}}%
\newcommand\fcp{{Fund.~Cosmic~Phys.}}%
\newcommand\gca{{Geochim.~Cosmochim.~Acta}}%
\newcommand\grl{{Geophys.~Res.~Lett.}}%
\newcommand\jcp{{J.~Chem.~Phys.}}%
\newcommand\jgr{{J.~Geophys.~Res.}}%
\newcommand\jqsrt{{J.~Quant.~Spec.~Radiat.~Transf.}}%
\newcommand\memsai{{Mem.~Soc.~Astron.~Italiana}}%
\newcommand\nphysa{{Nucl.~Phys.~A}}%
\newcommand\physrep{{Phys.~Rep.}}%
\newcommand\physscr{{Phys.~Scr}}%
\newcommand\planss{{Planet.~Space~Sci.}}%
\newcommand\procspie{{Proc.~SPIE}}%
\newcommand\helvet{{Helvetica~Phys.~Acta}}%


\pagerange{\pageref{firstpage}--\pageref{lastpage}} \pubyear{2009}

\maketitle

\label{firstpage}

\begin{abstract}
The time delay between the formation of the progenitor systems of Type Ia supernovae (SNe Ia) and their detonation is a vital discriminant between the various progenitor scenarios that have been proposed for them. We use SDSS optical and \textit{GALEX} ultraviolet observations of the early-type host galaxies of 21 nearby SNe Ia and quantify the presence or absence of any young stellar population to constrain the minimum time delay for each supernova. We find that early-type host galaxies lack `prompt' SNe Ia with time delays of $\lesssim$100 Myr and that $\sim$70\% SNe Ia have minimum time delays of 275 Myr -- 1.25 Gyr, with a median of 650 Myr, while at least 20\% SNe Ia have minimum time delays of at least 1 Gyr at 95\% confidence and two of these four SNe Ia are are likely older than 2 Gyr. The distribution of minimum time delays observed matches most closely the expectation for the single-degenerate channel with a main sequence donor. Furthermore, we do not find any evidence that sub-luminous SNe Ia are associated with long time delays.
\end{abstract}


\section{Introduction}
The progenitor channels of Type Ia supernovae (SNe Ia) are of vital importance both to our understanding of stellar evolution and to modern cosmology where SNe Ia are used as standard candles \citep{1998AJ....116.1009R, 1999ApJ...517..565P}. They are also a key part of our understanding of galaxy formation by virtue of their contribution to the energy budget and chemical evolution of their host galaxies \citep[e.g.][]{1983A&A...118..217G, 1986A&A...154..279M, 2004MNRAS.347..968P}. A range of recent observations have indirectly suggested that at  least some SNe Ia  can be produced by a variety of different progenitor systems, though these claims are tentative \citep{2003ApJ...582..915H, 2004Natur.431.1069R, 2007Sci...317..924P, 2008Natur.451..802V, 2009A&A...493.1081J}. Since no SN Ia progenitor system has been conclusively identified pre-explosion, we must use more indirect approaches to understand their origin.

The various progenitor scenarios proposed in the literature have very different predictions for the \textit{time delay} between the episode of star formation producing progenitor systems, and the time until the detonation of SNe Ia and so observational constraints on the distribution of time delays (DTD) are thus critical for discriminating between the various scenarios. We will discuss each of them in turn.

\subsection{Progenitor scenarios for SN Ia and their expected time delays}
Theory proposes that there are two main scenarios for the origin of SNe Ia; in both cases the system giving rise to the SN Ia is the thermonuclear explosion of a Carbon-Oxygen (CO) white dwarf (WD) in a binary system. We briefly describe the main channels here; for a full review of progenitor scenarios and explosion mechanisms, see \cite{2008NewAR..52..381P}.

\subsubsection{Single degenerate scenarios}
In the \textit{single degenerate scenario} (SD), the steady mass transfer from a main sequence (MS) donor star slowly builds the mass of a WD companion until it reaches approximately the Chandrasekhar mass\footnote{The point at which an explosive nuclear runaway occurs in a nonrotating CO WD may be slightly below the Chandrasekhar mass. \cite{1984ApJ...286..644N} calculate a value of $\sim1.378 ~\rm M_{\odot}$.}, initiating an SN Ia \citep{1991ApJ...367L..19N, 1999ApJ...522..487H, 2000A&A...362.1046L, 2004MNRAS.350.1301H, 2009A&A...493.1081J}. By the time of explosion, the donor can already be a slightly evolved subgiant. In this scenario, the time delay distribution is expected to peak of $\sim 670$ Myr with virtually no contribution past $\sim 1-1.5$ Gyr \citep{2004MNRAS.350.1301H}.

Both very short (or `prompt') and very long $\sim 1$ Gyr time delay for SD channels have been proposed: \cite{2008ApJ...679.1390H} suggest that massive (6-7 $M_{\odot}$) MS donors are viable, while \cite{2009MNRAS.395..847W} argue for Helium star donors. In 

both cases, the detonation is expected to occur within $\lesssim 100$ Myr. Very long SD time delays are possible if the donor is a low-mass red giant (RG)\citep{1996ApJ...470L..97H, 1999ApJ...522..487H}.

\subsubsection{Double degenerate scenarios}
In the \textit{double-degenerate channel} (DD) on the other hand, the thermonuclear explosion occurs when two white dwarfs in a binary system merge \citep{1984ApJS...54..335I, 1984ApJ...277..355W}
The range of possible time delays is set by the timescale for the formation of a WD binary system followed by the time needed for orbital decay via gravitational waves \citep{1983bhwd.book.....S}, which depends only on the binary separation following the common envelope phase.

\begin{figure*}
\begin{center}

\includegraphics[angle=90, width=0.9\textwidth]{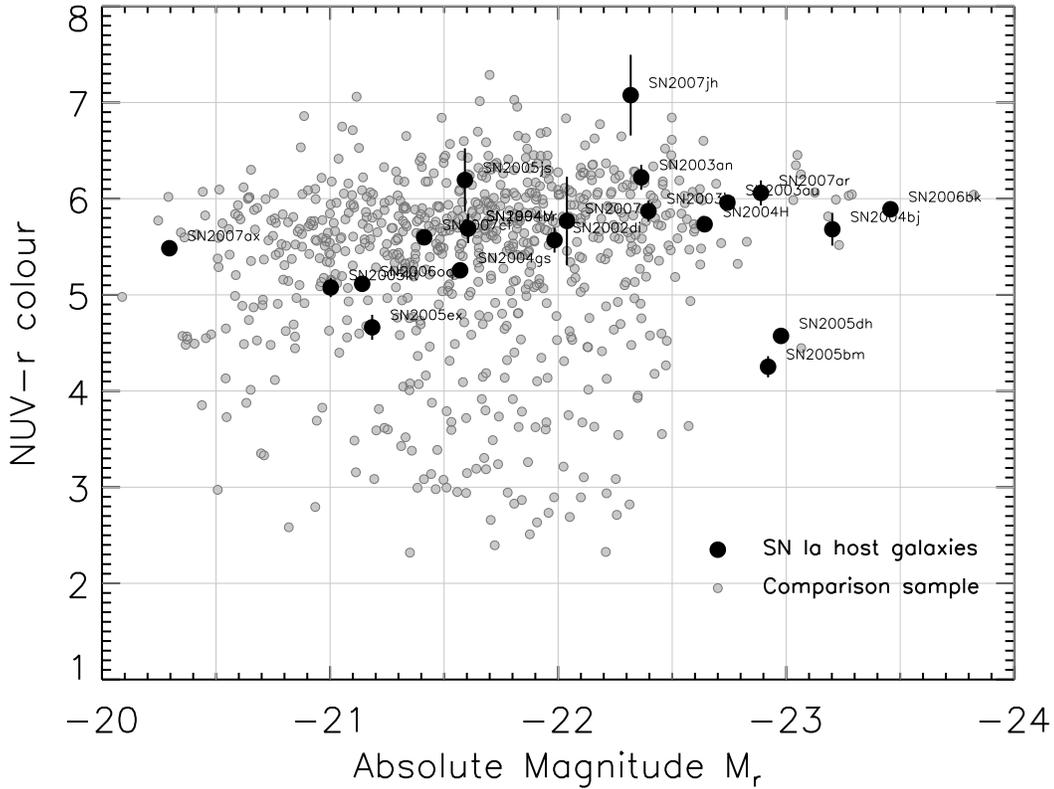}
\caption{The UV-optical colour-magnitude relation of a sample of normal early-type galaxies, and the SN Ia host galaxies examined in this work. The grey points are the comparison sample of early-type galaxies taken from \citet{2007ApJS..173..512S}, while the black points are the SN Ia host galaxies with \textit{GALEX} observations from \citet{2008MNRAS.388L..74F}. For the SN Ia host galaxies, we show the error bars on the $NUV-r$ colour; these are dominated by the error on NUV. We also label each host galaxy by the name of the supernova that occurred (c.f. Table \ref{tab:sample}).  }

\label{fig:uv_cmr}

\end{center}
\end{figure*}

\subsection{Early-type galaxies as an ideal laboratory}
Recent advances in our understanding of star formation -- and absence thereof -- in the local early-type population, together with a large volume of new observations make early-type galaxies an ideal laboratory for constraining the time delay distribution of low redshift SNe Ia. Local early-type galaxies have generally been considered to have formed at high redshift and devoid of any current or recent star formation \citep[e.g.][]{1992MNRAS.254..601B, 2005ApJ...621..673T}, implying that any SN Ia progenitor systems are old with ages of several Gyrs. This view has been revised in recent years by observations from the \textit{Galaxy Evolution Explorer} (\textit{GALEX}; \citealt{2005ApJ...619L...1M}) UV space satellite: \cite{2005ApJ...619L.111Y} found that a substantial fraction of early-type galaxies observed by \textit{GALEX} had UV-optical colours incompatible with any scenarios of enhanced extreme horizontal branch stars (UV upturn; for a review of the current state see \citealt{2008ASPC..392....3Y}) that could only be explained by the presence of small young stellar populations.

This discovery was quantified using UV-optical colours to determine that between 1/3 and half of the local luminous early-type galaxy population harboured young stellar populations with mass fractions of a few percent and ages up to approximately 1 Gyr \citep{2007ApJS..173..619K, 2006Natur.442..888S, 2007ApJS..173..512S} and their presence has been detected out to z $\sim 1$ \citep{2008MNRAS.388...67K}.

The inclusion of the \textit{GALEX} near- and far-UV broadband photometry (NUV, 1771-2831$\rm \AA$; FUV, 1344-1786$\rm \AA$; \citealt{2007ApJS..173..682M}) into the optical spectral energy distribution allows us to detect and quantify any young stellar populations in terms of age and mass-fraction, or to rule out their presence against the backdrop of an old, passively evolving bulk stellar population. This makes early-type galaxies in the local universe observed by \textit{GALEX} an ideal laboratory to constrain and quantify the the ages of SN Ia progenitors for those events which occurred in early-type galaxies. Late-type galaxies with extended star formation histories are not suitable for constraining time delays out to interesting ages, as they always harbour stars of very young ages from ongoing star formation.

\begin{table*}
\begin{center}
\caption{The sample of SNe Ia and their host galaxy properties}
\label{tab:sample}
\begin{tabular}{@{}llllllllll}
\hline
Galaxy Name & \multicolumn{1}{l}{Supernova} & \multicolumn{1}{l}{Redshift} & \multicolumn{1}{c}{$NUV-r$}  & \multicolumn{1}{c}{$M_{r}$} & \multicolumn{1}{l}{Best-fit}  & \multicolumn{1}{l}{Young age range} & \multicolumn{1}{c}{Youngest possible $t_{y}$} & \multicolumn{1}{c}{Youngest possible $t_{y}$}\\
&  \multicolumn{1}{l}{Name}& & & &  \multicolumn{1}{l}{young age $t_{y}$}&  &  \multicolumn{1}{l}{at 68\% confidence}   &  \multicolumn{1}{l}{at 95\% confidence}\\
& & & $mag$ & $mag$ & $Myr$ & $Myr$ & $Myr$ & $Myr$\\
\hline
\hline
   NGC4493 &        SN1994M  &     0.02316    &        5.69$\pm$        0.15 &      -21.61$\pm$       0.002   &        684  &    650   -  1980  &         719   &        558   \\
CGCG169-00 &       SN2002di  &     0.03639    &        5.57$\pm$        0.13 &      -21.98$\pm$       0.002   &        881  &    926   -  6670  &        1193   &        757   \\
    IC4258 &       SN2003an  &     0.03704    &        6.22$\pm$        0.13 &      -22.36$\pm$       0.002   & -&- &      6028   &       2550   \\
   NGC6095 &    SN2003au$^1$ &     0.03084    &        5.96$\pm$        0.07 &      -22.74$\pm$       0.002   &        587  &    531   -   650  &         558   &        480   \\
   NGC6109 &       SN2003ia  &     0.02954    &        5.87$\pm$        0.11 &      -22.40$\pm$       0.002   & -&- &      2424   &       1025   \\
CGCG044-04 &       SN2004bj  &     0.05015    &        5.68$\pm$        0.17 &      -23.20$\pm$       0.002   &        757  &    757   -  7764  &         926   &        650   \\
   NGC4493 &       SN2004br  &     0.02316    &        5.69$\pm$        0.15 &      -21.61$\pm$       0.002   &        684  &    650   -  1980  &         719   &        558   \\
CGCG089-05 &       SN2004gs  &     0.02664    &        5.26$\pm$        0.04 &      -21.57$\pm$       0.002   &        618  &    531   -   684  &         558   &        480   \\
    IC0708 &        SN2004H  &     0.03168    &        5.73$\pm$        0.05 &      -22.64$\pm$       0.002   & -&- &      4021   &       1255   \\
2MASXJ1520 &       SN2005bm  &     0.10350    &        4.25$\pm$        0.11 &      -22.92$\pm$       0.003   &        587  &    433   -   618  &         505   &        336   \\
   MRK0693 &       SN2005dh  &     0.03836    &        4.57$\pm$        0.03 &      -22.98$\pm$       0.002   &        618  &    531   -   837  &         587   &        480   \\
2MASXJ0141 &       SN2005ex  &     0.09000    &        4.66$\pm$        0.13 &      -21.18$\pm$       0.006   &        456  &    336   -   531  &         392   &        275   \\
2MASXJ0134 &       SN2005js  &     0.07968    &        6.19$\pm$        0.33 &      -21.59$\pm$       0.004   &        587  &    433   -   618  &         505   &        372   \\
2MASXJ0110 &       SN2005kt  &     0.06540    &        5.08$\pm$        0.10 &      -21.00$\pm$       0.004   &        684  &    650   -  1389  &         719   &        587   \\
CGCG193-01 &       SN2006bk  &     0.04953    &        5.89$\pm$        0.05 &      -23.46$\pm$       0.002   &        618  &    558   -   650  &         587   &        505   \\
   NGC3841 &       SN2006oq  &     0.02120    &        5.11$\pm$        0.04 &      -21.14$\pm$       0.002   &        837  &    719   -  1320  &         796   &        684   \\
CGCG294-03 &       SN2007ar  &     0.05293    &        6.06$\pm$        0.13 &      -22.89$\pm$       0.002   & -&- &      4449   &        796   \\
   NGC2577 &    SN2007ax$^2$ &     0.00686    &        5.49$\pm$        0.04 &      -20.30$\pm$       0.002   &        757  &    757   -  6028  &         837   &        650   \\
CGCG077-10 &    SN2007cf$^3$ &     0.03293    &        5.60$\pm$        0.07 &      -21.41$\pm$       0.002   & -&- &      1537   &        975   \\
2MASXJ1410 &       SN2007ei  &     0.10000    &        5.77$\pm$        0.46 &      -22.04$\pm$       0.005   & -&- &      1461   &        837   \\
CGCG391-01 &       SN2007jh  &     0.04080    &        7.08$\pm$        0.42 &      -22.32$\pm$       0.002   & -&- &      4021   &       2191   \\
\hline
\end{tabular}
\\
\begin{flushleft}
$^1$ Spectrum resembles SN1991bg \citep{2003IAUC.8085....2F}, potentially sub-luminous.\\
$^2$ Faintest SN Ia known \citep{2008ApJ...683L..29K}.\\
$^3$ Reported sub-luminous by \cite{2007CBET..958....1B}.\\
\end{flushleft}
\end{center}
\end{table*}

\subsection{Previous constraints on SN Ia time delays}

Previous studies have endeavoured to determine the distribution of time delays (DTD) of SNe Ia and connect them to progenitor scenarios. On the theory side, analytical calculations by \cite{2005A&A...441.1055G} and \cite{2005ApJ...629..915B} argue that the SD channel with short time delays is unlikely to account for all SNe Ia in passive systems. Similarly, \cite{2008ApJ...683L..25P} make the case that the SD channel alone can only account for the observed relationship between the SNe Ia rate and host galaxy star formation rate under unrealistic assumptions. Similarly, \cite{2004ApJ...613..200S, 2005ApJ...635.1370S} and \cite{2006MNRAS.368.1893F} attempted to re-construct the distribution of time delays from the cosmic star formation history, but they concluded that this approach is limited by uncertainties in the SFH. In addition, the small sample statistics of \cite{2004ApJ...613..200S, 2005ApJ...635.1370S} are a further limiting factor.

The properties of SNe Ia depend on the properties of their host galaxies. The brighter, slowly declining events tend to occur in star-forming host galaxies, while the dimmer, more rapidly declining SNe Ia are preferentially found in red, passive host galaxies \citep[see e.g.][]{2006ApJ...648..868S}. This has led to the hypothesis that the bright SNe Ia result from progenitor systems with short time delays in star-forming galaxies, while the dimmer events, hosted by galaxies without any current or recent star formation are due to scenarios with longer time delays.\cite{2006MNRAS.370..773M} argued for the existence of two separate SN Ia populations, a `prompt' component with a time delay of $\sim 100$ Myr, and a `delayed' component with time delays of 3--4 Gyr. \cite{2008A&A...479...49B} and \cite{2008PASJ...60.1327T} have analysed the host galaxies of SNe Ia and concluded that a substantial fraction of SNe Ia must have long time delays on the order of 2--3 Gyr. At the other end of the delay time range, \cite{2008A&A...492..631A} have argued that there is a sub-population of SNe Ia with very short time delays of less than 180 Myr.

\cite{2008ApJ...685..752G} recently argued to have found a direct link between the properties of individual SNe Ia and the bulk stellar populations of early-type host galaxies. They claim that that SNe Ia in host galaxies with older bulk ages and higher metallicities are dimmer and decay more rapidly. This has however been disputed by \cite{2009ApJ...691..661H} who find no such correlation. The tempting conclusion to draw here is that perhaps sub-luminous SNe Ia (sometimes called 1991bg-like, after the prototypical event) are from the long-time delay population, perhaps originating from a different explosion mechanism due to their progenitor channel \citep[e.g.][]{2001ApJ...554L.193H, 2008MNRAS.385...75T,2009ApJ...691..661H}. The consensus in the literature is that sub-luminous events are due to explosions with low $^{56}Ni$ masses \citep[e.g.][]{2002ApJ...568..791H, 2007ApJ...658..396F}.

Rather than infer time delays indirectly from the bulk stellar population via optical data, in this Paper we probe deeper into the star formation history of a sample of 21 SN Ia host galaxies and use \textit{GALEX} UV data to determine the presence or absence of minor episodes of recent star formation and thus constrain the minimum time delay for these SNe Ia.

This paper is organised as follows. In Section \ref{sec:sample}, we discuss the selection of early-type SN Ia host galaxies that have been observed by \textit{GALEX} and compare them to the general population. In Section \ref{sec:method}, we motivate and describe our SED analysis method and present the results from applying it in Section \ref{sec:results}, followed by a discussion in Section \ref{sec:discussion}.
We assume cosmological parameters $(\Omega_{m} = 0.3, \Omega_{\Lambda} = 0.7, H_{0} = 70)$, consistent with the \textit{Wilkinson Microwave Anisotropy Probe} (WMAP) Third Year results and their combination of results with other data \citep{2007ApJS..170..377S}.

\section{Sample Selection}
\label{sec:sample}

\subsection{SN Ia early-type host galaxies}
We take the catalogue of visually inspected early-type galaxies that have hosted an SN Ia presented by
\cite{2008MNRAS.388L..74F}. The objects in this sample were culled by visually inspecting all host galaxies of SN Ia from the CfA list of supernovae\footnote{\texttt{http://www.cfa.harvard.edu/iau/lists/Supernovae.html}} that overlap with the Sloan Digital Sky Survey DR6 (SDSS; \citealt{2008ApJS..175..297A}). The visual classification is vital to avoid the inclusion of systems with weak disks or spiral arms that can contaminate selections by proxies such as structural parameters or optical colours \citep[see e.g.][]{2007MNRAS.382.1415S, 2008MNRAS.389.1179L}. We then limit this to those objects with observations in the \textit{GALEX} GR4 archive\footnote{\texttt{http://galex.stsci.edu/GR4/}}, regardless of the survey; in total, this leaves us with a sample of 21 objects (see Table \ref{tab:sample}). We did not require a \textit{GALEX} detection, only an observation, to avoid a bias towards UV-bright objects; however, all objects in our sample have a detection in at least one \textit{GALEX} filter.

 Where \textit{GALEX} data at multiple depths was available, the deeper image was used. The optical SDSS host galaxy and the corresponding \textit{GALEX} detection were matched by hand.

\begin{figure}
\begin{center}

\includegraphics[angle=90, width=00.49\textwidth]{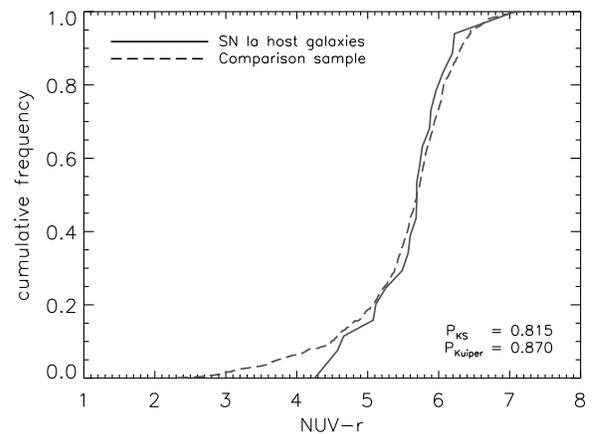}
\caption{The cumulative $NUV-r$ colour distribution of both the SN Ia host galaxies (\textit{solid line}) and the comparison sample (\textit{dashed line}). The two colour distributions are closely matched. There appears to be a lack of very blue SN Ia hosts ($NUV-r < 4$), but neither a Kolmogorov--Smirnov test nor a Kuiper test indicate that the two populations are drawn from a different parent distribution; the lack of very blue SN Ia hosts is thus not statistically significant.}

\label{fig:nuv_r_cumu}

\end{center}
\end{figure}

\subsection{Are SN Ia host galaxies different from normal early-type galaxies}
Before we begin our quantitative analysis of the recent star formation histories of the SN Ia host galaxies, we compare their UV-optical properties to those of normal early-type galaxies. For comparison, we use the sample of early-type galaxies with \textit{GALEX} observations presented in \cite{2007ApJS..173..512S}. This sample is magnitude-limited ($r < 16.8$), limited to $0.05 < z < 0.1$ and visually inspected for early-type morphology.  We must preface this comparison by pointing out that while this comparison sample is well-defined, the SN Ia host galaxies sample is not; it is drawn from a heterogeneous parent sample of SN Ia composed of various surveys, all with their own selection effects and serendipitous discoveries of individual supernovae. Keeping this caveat in mind, we proceed to compare the two.

In Figure \ref{fig:uv_cmr}, we show the $NUV-r$ colour-magnitude diagram for the SN Ia host galaxies (black points) and the comparison sample from \citet{2007ApJS..173..512S} (grey points). The comparison sample reasonably covers the parameter space of the SN Ia host galaxies. The host galaxies of SN2005dh and SN2005bm are both very blue in $NUV-r$ and very luminous. We compare the $NUV-r$ colour distributions of the host galaxies and the comparison sample in Figure \ref{fig:nuv_r_cumu}. While the cumulative distribution of the host galaxies might appear to indicate a lack of host galaxies at the bluest $NUV-r$ colour, both a Kolmogorov--Smirnov and a Kuiper test indicate that the two distributions are consistent with being drawn from the same parent distribution, and so this lack of very blue SN Ia hosts is not statistically significant at the 95\% level.

\section{Method}
\label{sec:method}

\subsection{Quantitative analysis of the recent star formation history}
\begin{figure*}
\begin{center}

\includegraphics[angle=90, width=00.49\textwidth]{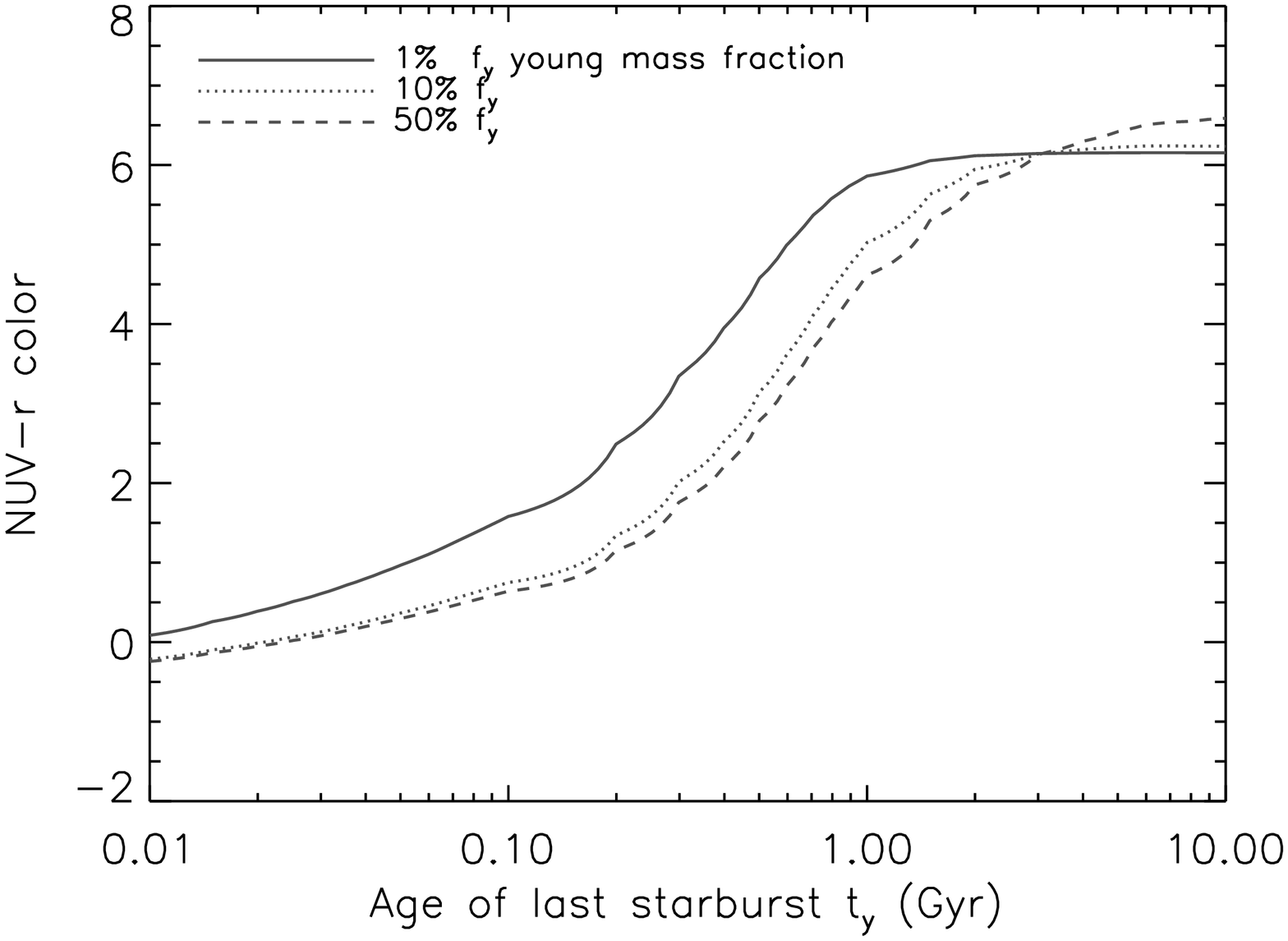}
\includegraphics[angle=90, width=00.49\textwidth]{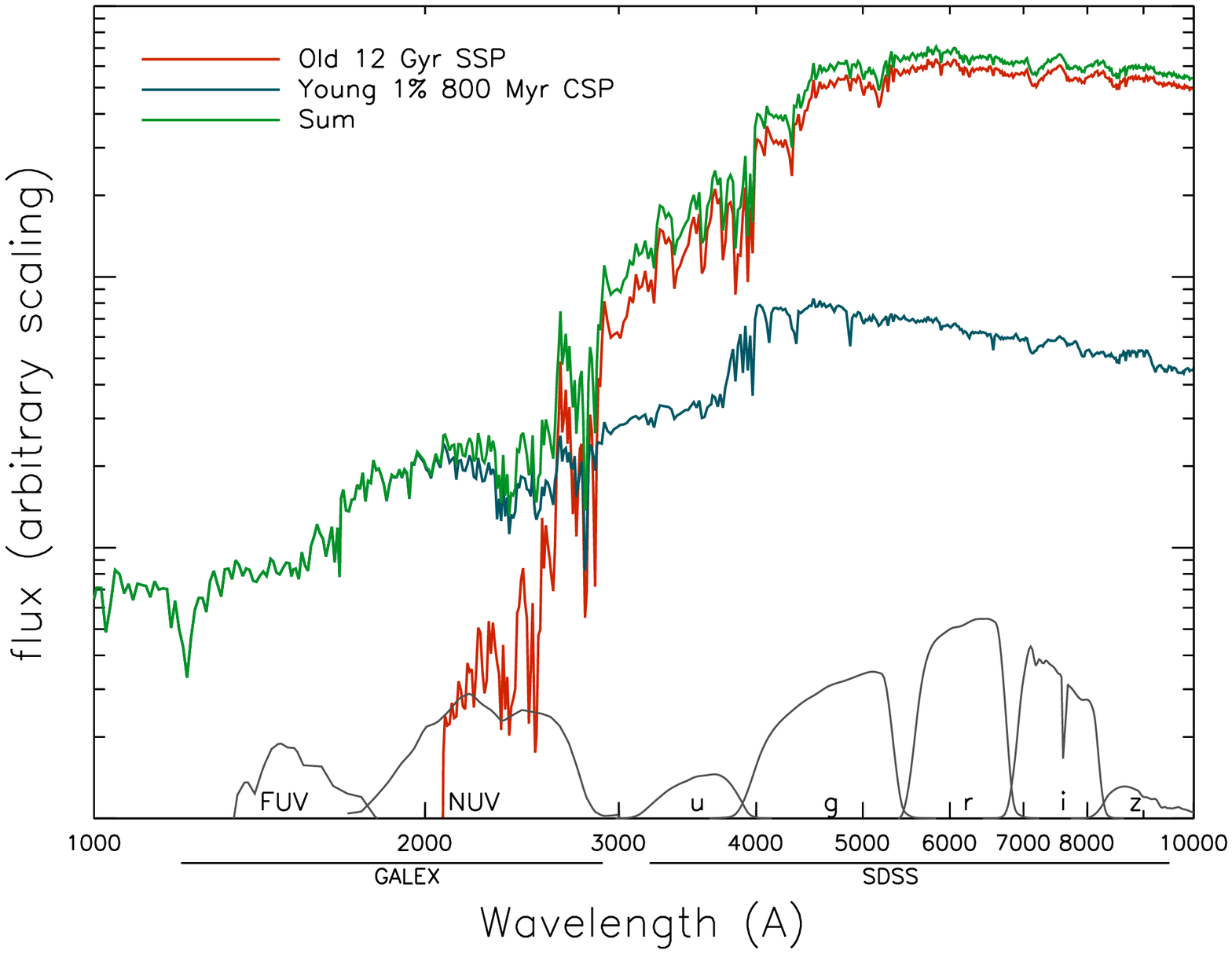}
\caption{The utility of \textit{GALEX} UV data. \textit{Left:} The evolution of $NUV-r$ as a function of time elapsed since the last starburst on top of a 12 Gyr old passively evolving SSP. We show this curve for three different mass-fractions $f_{y}$. All three models are extinguished by $E(B-V) = 0.1$. This demonstrates the enormous range in $NUV-r$ colour for even small young populations and its strong dependence on the young age, and its weak dependence on the mass fraction. \textit{Right:} A sum of a 12 Gyr old \citet{2005MNRAS.362..799M} single burst (\textit{red}) and a 1\% mass-fraction 800 Myr old CSP  (\textit{blue}) summing to a total SED  (\textit{green}). In the optical SDSS filters (indicated by their filter transmission curves), the addition of this small, young burst hardly changes the SED, while the difference becomes apparent and dominates in the UV filters probed by \textit{GALEX}.}

\label{fig:color_evolution}

\end{center}
\end{figure*}

The exquisite sensitivity of the near-UV to small amounts of young stellar populations makes it the ideal tool to quantify any young populations, or to rule out their presence down to very small levels. The old bulk stellar populations of massive early-type galaxies makes the exercise of measuring small young components easier. In Figure \ref{fig:color_evolution} (left), we show the evolution of the $NUV-r$ colour as a function of the age of the young component for a range of parameters to illustrate the strong age sensitivity of $NUV-r$. On the right, we show a composite SED of a single 12 Gyr old burst (red) and a 800 Myr young burst of a 1\% mass fraction (blue). While it hardly changes the total SED (green) in the optical wavelengths probed by SDSS, the young population dominates the UV wavelengths probed by \textit{GALEX}.

We parameterise the star formation history as two components: a burst on top of a old component. This parameterisation has been successfully implemented before in order to answer questions of galaxy formation \citep{2000ApJ...541L..37F, 2007ApJS..173..619K, 2007MNRAS.382.1415S, 2009ApJ...690.1672S}. Any young stellar population, even if small,  is luminous compared to the underlying older population. Therefore, by marginalising over all possible star formation histories before the most recent episode of star formation, we are able to constrain it, usually out to ages of $\sim1$ Gyr and rule out any such episodes out to 1--2 Gyr.

The properties of any young component is degenerate with a number of other parameters. Factors include the metallicity $Z$, which must be varied over the entire plausible range of metallicities, and should include an internal metallicity distribution with a tail ranging to very low metallicities. For old ages, both the low metallicity tail \citep{1997ApJ...476...28P, 2000ApJ...541..126M} and high metallicities can contribute moderate UV flux \citep{1997ApJ...482..677Y, 1997ApJ...486..201Y}. Since we do not wish to erroneously attribute any weak UV flux to young stars, we must account for this alternate, old origin. The availability of far-UV data from the \textit{GALEX} $FUV$ filter aids somewhat in the differentiation, as these old populations tend to be hotter than young populations of a few hundred Myr and so have a steeper rising UV spectrum. We model our internal metallicity distribution on the observed distribution of the bulges of nearby galaxies \citep{2005AJ....130.1627S, 2000AJ....120.2423H, 2002AJ....123.3108H, 1999AJ....117..855H}. We range the metallicity $Z$ such that the mass-weighted total metallicity ranges between $\frac{1}{20} Z_{\odot}- 3.25 Z_{\odot} $.

\begin{table}
\begin{center}
\caption{Model Parameters}
\label{tab:parameters}
\begin{tabular}{@{}lll}
\hline
\multicolumn{1}{l}{Parameter}&  & \multicolumn{1}{l}{Range}\\
&   \\
\hline
\hline
Old age 			& $t_{o}$	&   1--15 Gyr \\
Young age$^1$		& $t_{y}$ 	& 0.1--15 Gyr\\
Metallicity		& $Z$ 		& $\frac{1}{20}- 3.25 Z_{\odot} $ \\
Dust extinction$^2$ 	&$E(B-V)$ 	& 0--0.3\\
Young mass fraction 	& $f_{y}$ 	& 1--100\%\\
\hline
\end{tabular}
\\
\begin{flushleft}
$^1$ The young age $t_{y}$ is always restricted to be less than the old age $t_{o}$.\\
$^2$ We use a \cite{2000ApJ...533..682C} extinction law.\\
\end{flushleft}
\end{center}
\end{table}

\begin{figure*}
\begin{center}

\includegraphics[angle=90, width=\textwidth]{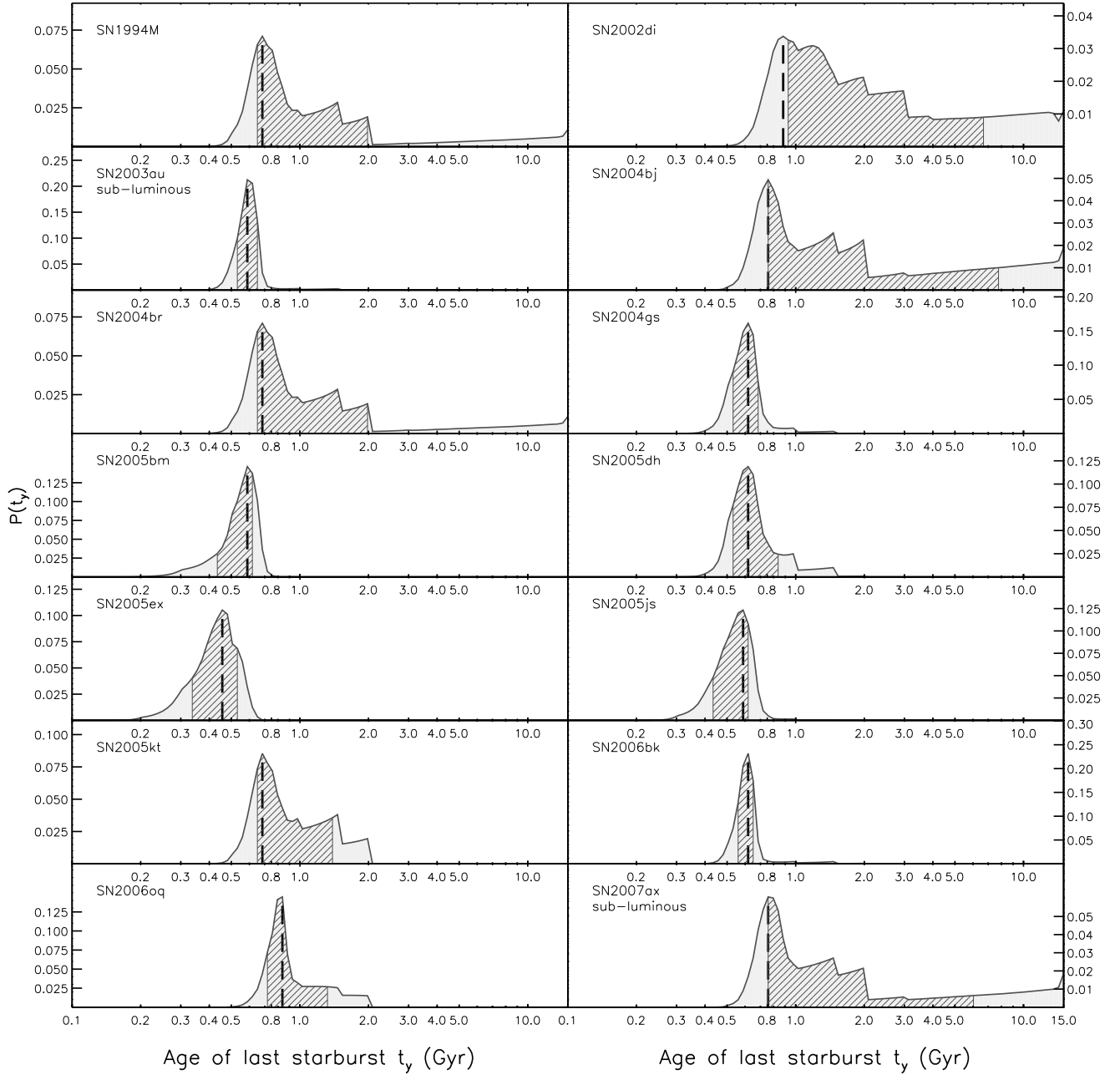}
\caption{The probability distribution functions $P(t_{y})$ characterising the age of the most recent episode of star formation for those SN Ia host galaxies in our sample where a young component is resolved. We label each panel by the name of the supernova and its host galaxy. Note that SN1994M and SN2004br both occurred in NGC4493, so the $P(t_{y})$ is identical. The black dashed line indicates the best-fit age, while the single-hatched region shows the two-sided 68\% confidence interval.}

\label{fig:prob_young}

\end{center}
\end{figure*}

\begin{figure*}
\begin{center}

\includegraphics[angle=90, width=\textwidth]{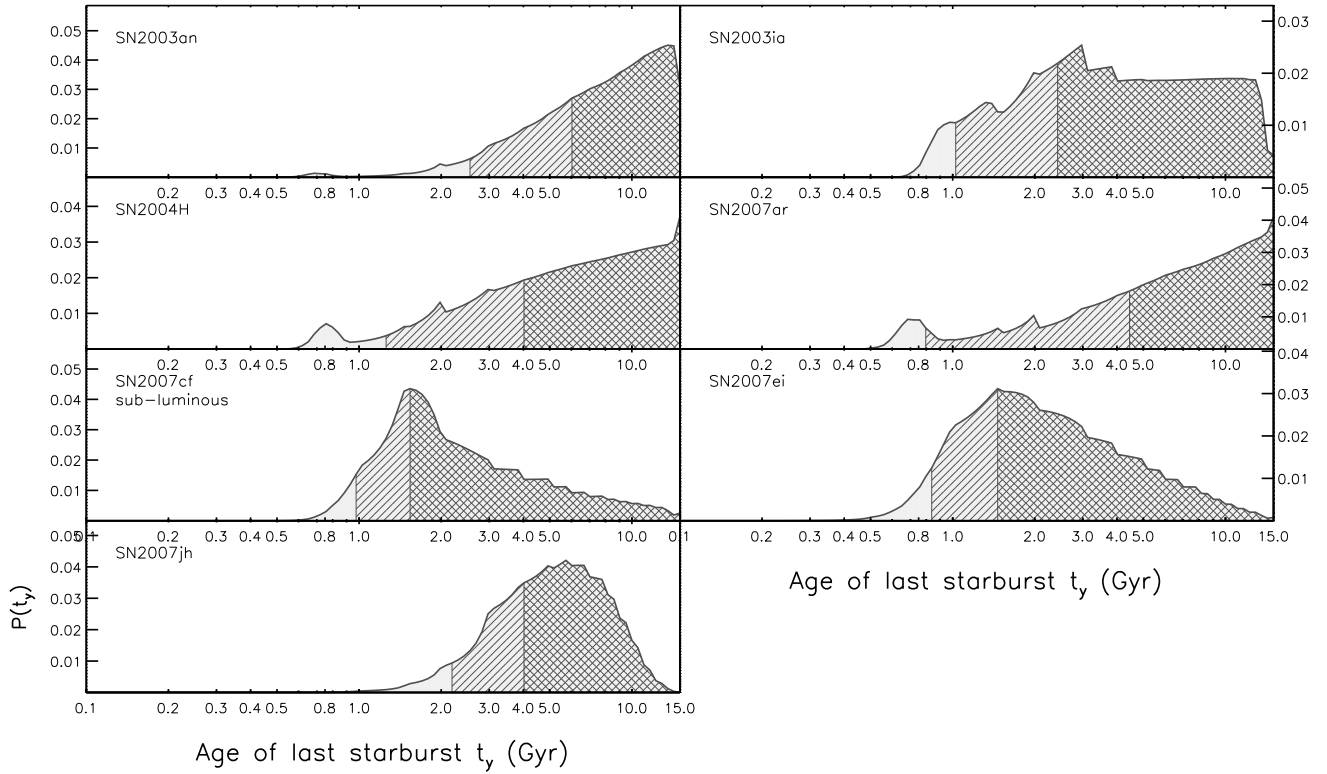}
\caption{The probability distribution functions $P(t_{y})$ characterising the age of the most recent episode of star formation for those SN Ia host galaxies in our sample where no young component is resolved. As in Figure \ref{fig:prob_young}, we label each panel by the name of the supernova and its host galaxy. The single-hatched region shows the one-sided 95\% confidence limit, while the double-hatched region shows the 68\% limit. }

\label{fig:prob_old}

\end{center}
\end{figure*}

Further factors are extinction by dust, which we model using the law of \cite{2000ApJ...533..682C} and range the colour excess $E(B-V)$ from 0--0.3 to account for levels of extinction typical in early-type galaxies\footnote{The measured range of $E(B-V)$ for passively evolving early-types is very low; see e.g. Figure 11 in \cite{2007MNRAS.382.1415S}.}, and the age of the bulk old stellar population. We model this dominant component as a single stellar population (SSP) of varying age from 1--15 Gyr. The young component is represented by a composite stellar population describing a brief burst of star formation declining with an e-folding time of 100 Myr. The key parameters of this young burst are its age $t_{y}$ and its mass-fraction $f_{y}$, which are somewhat degenerate with each other. \cite{2007MNRAS.382.1415S} showed that the folding in of the luminosity-weighted age and metallicity measured from stellar absorption indices (Lick indices) can break this degeneracy, but SDSS spectra are not available for all SN Ia host galaxies, and the majority of our host galaxies are so nearby that the $3\arcsec$ spectroscopic fibre of SDSS samples only the very central part\footnote{In one case, the SDSS spectroscopic target selection algorithm selected two off-nucleus positions.}, which represents a biased sub-sample of the bulk stellar population, as many early-type galaxies exhibit radial gradients in their stellar population parameters \citep[][]{1993MNRAS.262..650D, 1993MNRAS.265..553C, 2000AJ....119.1645T}.

The degeneracy between the mass-fraction and the age of any young stellar population represents a fundamental limit to the constraints that can be derived from applying this method. The limit on the age of the young population $t_{y}$ that we can derive is a function of the lowest mass-fraction $f_{y}$ that we probe down to. The age $t_{y}$ is not a strong function of the mass-fraction $f_{y}$ until below the $\sim1\%$ level, where even UV observations fail to be sufficiently sensitive. We thus range in mass-fraction $f_{y}$ from 100\% down to 1\%, far lower than is possible with optical photometry alone. All constraints on the age $t_{y}$ derived here are thus subject to the caveat that we do not account for young populations below the $1\%$ level.

We generate a library of photometric data points as a function of the parameters described above, resulting in 6.5 million model SEDs in the restframe of each SN Ia host galaxy using the models of \cite{2005MNRAS.362..799M, 1998MNRAS.300..872M}. We fit each host galaxy to all library SEDs and compute the $\chi^2$ statistic for each, convert them to probabilities, and marginalise over all parameters except for the young age $t_{y}$ in order to obtain the probability distribution function $P(t_{y})$. In order to account for unknown systematic errors in the SDSS and \textit{GALEX} zeropoints and other effects, we add in quadrature a uniform error of 0.1 mag to each band.

We have tested the effect of adding various amounts of systematic error to broad-band photometry for similar codes in general, and for the data set and specific implementation
used in this paper. By increasing the amount of assumed systematic error, the minimum set of parameters ($t_y$, $f_y$) do not change and individual fits do not change from cases
where a young population is detected to one where a young population is ruled out. Increasing the assumed error increases the size of the errors on ($t_y$, $f_y$) at the minimum,
or lowers the limit on $t_y$ in cases where no young component is detected. While the SDSS photometric system is exquisitely calibrated, down to the 1-3\% level for DR6\footnote{See \texttt{http://www.sdss.org/dr6/} and \cite{2008ApJS..175..297A}}, the photometric repeatability of \textit{GALEX} is $\sim$0.03 and 0.05 mag in NUV and FUV, respectively \citep{2007ApJS..173..682M}. Given this, we believe a 0.1 mag error may be a conservative overestimate. We note that the main driver in detecting and ruling out the presence of young stellar populations are UV-optical colours whose observed range spans 5-7 mag, substantially larger than the 0.1 mag error.

\section{Results}
\label{sec:results}

We present the results of our SED fitting analysis in Figures \ref{fig:prob_young} and \ref{fig:prob_old} and Table \ref{tab:sample}. For each supernova and host galaxy in our sample, we plot the marginalised probability distribution function $P(t_{y})$ from 100 Myr to 15 Gyr. The supernova host galaxies in our sample can be divided into two classes: those where a small young stellar populations with ages of $\lesssim 1$ Gyr, are present, and those where any such young population is ruled out down to 1\% by mass-fraction. This limit corresponds to the age resolution of the near-UV (c.f. Figure \ref{fig:color_evolution}). For those where a young component is resolved, we can calculate the best-fit age (Figure \ref{fig:prob_young}), while for those without, we can only report lower limits on the age of the youngest stellar population -- and therefore SN Ia progenitor -- present (Figure \ref{fig:prob_old}). For comparison, the results for the general early-type population are presented in \cite{2007ApJS..173..619K}.

\subsection{SNe Ia host galaxies with detected young stellar populations}
For 14 out of 21 SN Ia host galaxies in our sample, we do detect and constrain the presence of a small (few percent) young stellar population in the host galaxy. The age of this young population represents the shortest possible time delay for these SNe Ia, though it is of course possible that these particular SNe have \textit{longer} time delays, up to the age of the bulk stellar population. In Figure \ref{fig:prob_young}, we show $P(t_{y})$ for all 14 host galaxies with resolved young components\footnote{We note that for three of these -- SN2002di, SN2004bj and 2007ax -- there is a substantial tail to ages of several Gyr and the best-fit age is around 1 Gyr; these host galaxies straddle the limit of ages that are detectable.}. In each panel, we indicate the  best-fit age with a dashed black line. We compute the 68\%  two-sided confidence interval and shade it with single gray hatches. In Table \ref{tab:sample}, we report the best-fit ages and the two-sided confidence interval.

\subsection{SNe Ia with no detected young stellar populations}
For the remaining 7 SN Ia host galaxies, we do not detect the presence of any young stellar population down to the 1\% level in mass-fraction. For these, it only makes sense to compute a one-sided confidence interval. We shade the 68\% and 95\% confidence limits on the $P(t_{y})$ distribution in Figure \ref{fig:prob_old} and report these limits in Table \ref{tab:sample}. We also give these one-sided confidence levels for those host galaxies \textit{with} detected young components.

At 95\% confidence, we find that four SNe Ia (SN2003an, SN2003ia, SN2004H and SN2007jh) have time delays of at least 1 Gyr, with SN2003an and SN2007jh with minimum time delays in excess of 2 Gyr. Given the caveats discussed in Section \ref{sec:method}, \textit{this implies that long time delays for SNe Ia do occur in nature and at least $\sim 20\%$ (4/21) of SNe Ia in early-type host galaxies have time delays longer than 1 Gyr.} Due to the inclusion of \textit{GALEX} observations, the robustness of this statement is greater than that of any previous attempts to establish long time delays for some SNe Ia.

\section{Discussion}
\label{sec:discussion}

\subsection{The distribution of minimum time delays for SN Ia}
We are now able to plot the distribution of \textit{minimum} time delays (DmTD) for SNe Ia that occurred in early-type host galaxies. In Figure \ref{fig:age_hist}, we plot the histograms of the DmTD for SNe Ia at the 95\% limit. The minimum age distribution ranges from 275 Myr to 1.25 Gyr, with two SNe at $\sim 2$ Gyr and has a mean age of 650 Myr. There are \textit{no} cases where ages approaching $\sim 100$ Myr or below are allowed, so it appears that there are no `prompt' SNe Ia in our sample\footnote{The definition of what constitutes a `prompt' time delay varies somewhat in the literature. While a value of $\sim 100$ Myr is common, others are also used. For example \cite{2006ApJ...648..868S} use 500 Myr. By their definition, a number of the SNe Ia studies here would be classified as prompt.}.

\subsection{Sub-luminous SNe Ia?}
Several works have remarked that sub-luminous SNe Ia occur in early-type galaxies \citep[e.g.][]{2001ApJ...554L.193H}. Under the assumption that these host galaxies contain only old stellar population, this was taken to be evidence that sub-luminous SNe Ia have very long time delays on the order of several Gyr \citep{2001ApJ...554L.193H}. Although our sample of sub-luminous SNe Ia is very small, we find that in the case of two sub-luminous SN Ia, a shorter time delay is possible. Our sample contains three sub-luminous SNe Ia, SN 2003au, SN2007ax and SN2007cf. Of these, SN2003au and SN2007ax (the faintest SN Ia observed yet; \citealt{2008ApJ...683L..29K}) both are in host galaxies with young stellar components, while only SN 2007cf has a 95\% lower confidence limit of $\sim 1$ Gyr, i.e. only one sub-luminous SN Ia definitely has a long time delay. While this does \textit{not} prove that sub-luminous SNe Ia have short time delays, it clearly leaves open the possibility that they do.

\subsection{Implications for progenitor channels}
What do our results imply for the various proposed progenitor channels? There are two ways we can explore implications of the results in this paper for progenitor channels. We can assume that the bulk of SNe Ia originate from the youngest stellar population present in the host galaxy, rather than from the old bulk population. This assumption implies that the DmTD in Figure \ref{fig:age_hist} is a good approximation of the true DTD and so we can  take the DmTD and compare them to the predictions of various theoretical calculations of delay times. We stress that we cannot directly test with the current data whether this assumption is a good one and it remains plausible that the DmTD does not approximate the DTD well at all and that the young populations detected by \textit{GALEX} are mostly unrelated to the observed SNe Ia. In the following discussion, we explore what the DmTD implies for progenitor channels of SNe Ia assuming that it closely approximates the DTD:

\begin{figure}
\begin{center}

\includegraphics[angle=90, width=0.49\textwidth]{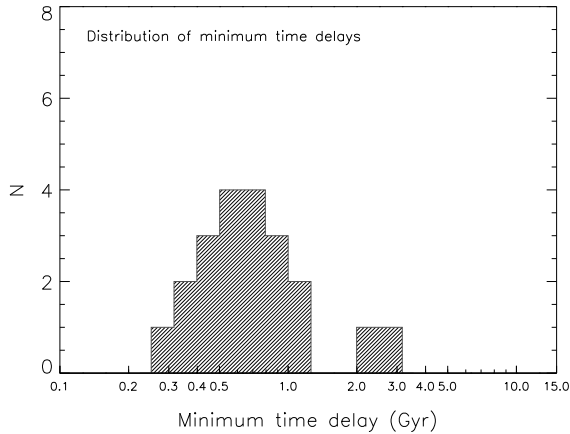}
\caption{The distribution of minimum time delays for the SNe Ia in our sample. These limits are the one-sided confidence limits derived from the $P(t_{y})$ distributions in Figures \ref{fig:prob_young} and \ref{fig:prob_old} (also given in Table \ref{tab:sample}). The mean and median age are 800Myr and 650 Myr respectively. }

\label{fig:age_hist}

\end{center}
\end{figure}

\subsubsection{Short time delay scenarios ($\lesssim100$ Myr)}
Scenarios that give rise to `prompt' SNe Ia, such as massive MS or Helium star donors \citep{2008ApJ...679.1390H, 2009MNRAS.395..847W} do not appear to occur in our sample. This is not necessarily obvious, as a few percent of early-type galaxies do have optically blue colours and host sufficiently young stellar populations \citep{2007MNRAS.382.1415S}, though the sample used in this Paper does not contain any such objects.

\subsubsection{Intermediate time delay scenarios ($100-1000$ Myr)}
The majority of SNe Ia in early-types have minimum time delays on the order of a few hundred Myr. If we assume that the actual DTD is approximated well by Figure \ref{fig:age_hist}, then the binary population synthesis predictions for the SD MS + CO WD scenario matches the distribution of ages well \citep[see][]{2004MNRAS.350.1301H}. 

\subsubsection{Long time delay scenario ($\gtrsim1$ Gyr)}
Regardless of whether Figure \ref{fig:age_hist} is a good representation of the actual DTD, a minority of SNe Ia must have long time delays in excess of 1 Gyr. The two progenitor channels that might account for these long time delays are the DD channel, can naturally produce such long time delays \citep{1984ApJS...54..335I, 1984ApJ...277..355W}, and the SD channel with RG donors. Calculations for the rate due to WD+RG progenitor systems predict that they cannot account for all SNe Ia in passive host galaxies \citep[e.g.][]{1998ApJ...497..168Y,2004MNRAS.350.1301H}

\section{Summary}
We have used SDSS optical and \textit{GALEX} UV photometry to measure the minimum time delay for a sample of 21 SNe Ia hosted in local early-type galaxies. We constrain the age, or rule out the presence, of any young stellar population in the host galaxies down to the 1\% level by mass. From this, we are able to construct the distribution of minimum time delays (DmTD) for the 21 SNe Ia in our sample. We find:

\begin{enumerate}
\item There are no prompt ($\lesssim 100$ Myr) SNe Ia in our sample.

\item For 14 out of 21 SN Ia host galaxies, we detect and constrain young stellar populations, yielding a range of minimum time delays of 275 Myr to 1.25 Gyr, with a mean of 650 Myr.

\item For 4 out of 21 SN Ia host galaxies, we can rule out the presence of any young stellar populations younger than 1 Gyr, implying minimum time delays longer than that. For two of these, the 95\% lower confidence limits rule out any time delays shorter than 2 Gyr.

\item Two out of three sub-luminous SNe Ia occurred in host galaxies with detected (few 100 Myr) young populations. Only one occurred in a host galaxy where young populations ($\lesssim 1$ Gyr) can be ruled out.
\end{enumerate}

We find that the predictions for the single degenerate MS + CO WD channel \citep{2004MNRAS.350.1301H} best matches the distribution of minimum time delays for the majority of our sample and that at least four require a scenario allowing for time delays greater than 1 Gyr, such as a RG + CO WD or DD channel. Under the extreme assumption that the DmTD is a good approximation of the true DTD, the SD MS + CO WD channel could account for up to 70\% of all SNe Ia in early-type galaxies, though the actual percentage could be much lower in case this assumption is not warranted.

\section*{Acknowledgements}
I am grateful for many comments and discussions to Stephen, Justham, Sugata Kaviraj, Zhanwen Han, Philipp Podsiadlowski, Brooke Simmons, Shanil Virani, Sukyoung K. Yi and Christian Wolf. I thank the anonymous referee for numerous comments and suggestions that have improved this work.

\textit{GALEX} (Galaxy Evolution Explorer) is a NASA Small Explorer, launched in 2003 April. We gratefully acknowledge NASA's support for construction, operation, and science analysis for the GALEX mission, developed in cooperation with the Centre National d'Etudes Spatiales of France and the Korean Ministry of Science and Technology.

Funding for the SDSS and SDSS-II has been provided by the Alfred P. Sloan Foundation, the Participating Institutions, the National Science Foundation, the U.S. Department of Energy, the National Aeronautics and Space Administration, the Japanese Monbukagakusho, the Max Planck Society, and the Higher Education Funding Council for England. The SDSS Web Site is http://www.sdss.org/.

The SDSS is managed by the Astrophysical Research Consortium for the Participating Institutions. The Participating Institutions are the American Museum of Natural History, Astrophysical Institute Potsdam, University of Basel, University of Cambridge, Case Western Reserve University, University of Chicago, Drexel University, Fermilab, the Institute for Advanced Study, the Japan Participation Group, Johns Hopkins University, the Joint Institute for Nuclear Astrophysics, the Kavli Institute for Particle Astrophysics and Cosmology, the Korean Scientist Group, the Chinese Academy of Sciences (LAMOST), Los Alamos National Laboratory, the Max-Planck-Institute for Astronomy (MPIA), the Max-Planck-Institute for Astrophysics (MPA), New Mexico State University, Ohio State University, University of Pittsburgh, University of Portsmouth, Princeton University, the United States Naval Observatory, and the University of Washington.

This research has made use of NASA's Astrophysics Data System Bibliographic Services.

\bibliographystyle{mn}

\bsp

\label{lastpage}

\end{document}